\documentclass[aps,prl,showpacs,floatfix,twocolumn,amsmath,amssymb,preprintnumbers]{revtex4-1}
\usepackage{mathrsfs}
\usepackage[figuresright]{rotating}
\usepackage{amsmath}
\usepackage{amssymb}
\usepackage{graphicx}
\usepackage{color}
\usepackage{dcolumn}
\usepackage{bm}
\usepackage[breaklinks=true,colorlinks=true,linkcolor=blue,urlcolor=blue,citecolor=blue]{hyperref}
\UseRawInputEncoding   

\usepackage{soul}

\makeatletter

\newcommand{\Rmnum}[1]{\expandafter\@slowromancap\romannumeral #1@}
\makeatother

\begin{document}
\author{Kangqiao Cheng$^{1\color{blue}{*}}$}
\author{Shuo Zou$^{1}$}
\email[]{These authors contribute equally to this work.}
\author{Huanpeng Bu$^{2}$}
\author{Jiawen Zhang$^{3}$}
\author{Shijie Song$^{3}$}
\author{Hanjie Guo$^{2}$}
\email[]{hjguo@sslab.org.cn}
\author{Huiqiu Yuan$^{3}$}
\email[]{hqyuan@zju.edu.cn}
\author{Yongkang Luo$^{1}$}
\email[]{mpzslyk@gmail.com}
\address{$^1$Wuhan National High Magnetic Field Center and School of Physics, Huazhong University of Science and Technology, Wuhan 430074, China;}
\address{$^2$Songshan Lake Materials Laboratory, Dongguan, Guangdong 523808, China;}
\address{$^3$ Center for Correlated Matter and School of Physics, Zhejiang University, Hangzhou 310058, China.}

\date{\today}

\title{Synthesis and physical properties of Ce$_2$Rh$_{3+\delta}$Sb$_4$ single crystals}

\begin{abstract}

Millimeter-sized Ce$_2$Rh$_{3+\delta}$Sb$_4$ ($\delta\approx 1/8$) single crystals were synthesized by a Bi-flux method and their physical properties were studied by a combination of electrical transport, magnetic and thermodynamic measurements. The resistivity anisotropy $\rho_{a,b}/\rho_{c}\sim2$, manifesting a quasi-one-dimensional electronic character. Magnetic susceptibility measurements confirm $\mathbf{ab}$ as the magnetic easy plane. A long-range antiferromagnetic transition occurs at $T_N=1.4$ K, while clear short-range ordering can be detected well above $T_N$. The low ordering temperature is ascribed to the large Ce-Ce distance as well as the geometric frustration. Kondo scale is estimated to be about 2.4 K, comparable to the strength of magnetic exchange. Ce$_2$Rh$_{3+\delta}$Sb$_4$, therefore, represents a rare example of dense Kondo lattice whose Ruderman-Kittel-Kasuya-Yosida exchange and Kondo coupling are both weak but competing.

\end{abstract}


\maketitle

\section{\Rmnum{1}. Introduction}

Heavy-fermion compounds are prototypical strongly correlated systems whose electrons can behave rather unconventionally. Examples of the latter include the ubiquitous strange-metal and/or non-Fermi-liquid behavior observed in the vicinity of quantum critical points (QCPs) characterized by sub-quadratic temperature dependent resistivity ($\rho \propto T^n$, where $n$ substantially less than 2), divergent electronic specific-heat coefficient (e.g. $C/T \propto -\log T$), etc\cite{Stewart-RMPNFL,Phillips-SM}. QCP refers to a continuous termination of a phase transition that leads to the point at absolute zero separating the quantum-ordered and -disordered states on the phase diagram. Of particular interest are ferromagnetic (FM) QCPs. An earlier `common sense' stated that FM QCPs are usually avoided in clean systems by the occurrence of a 1st-order transition\cite{Brando-FMQCP,Pfleiderer-1st-order}, the intersection of antiferromagnetic (AFM) phases\cite{Sullow-CeRu2Ge2,LuoY-CeFeAsPO,Jesche-CeFePO2017}, or a Kondo cluster glass phase\cite{Westerkamp-Clusterglass,Lausberg-CeFePOAvoid}. However, the past decade has witnessed the discovery of a handful of counterexamples, YbNi$_4$(P$_{1-x}$As$_x$)$_2$\cite{Krellner-YbNi4P2FMQCP,Steppke-YbNi4P2} and CeRh$_6$Ge$_4$\cite{Shen-CeRh6Ge4FMQCP}, where FM QCP seems realized. Irrespective of the inevitable disorder effect caused by chemical doping in YbNi$_4$(P$_{1-x}$As$_x$)$_2$, the pressure-induced continuous suppression of order parameter and the emergent strange-metal behavior in the pure ferromagnet CeRh$_6$Ge$_4$ provide a pristine paradigm of FM QCP\cite{Shen-CeRh6Ge4FMQCP}. The key common features of these two compounds are: (i) they both contain quasi-one-dimensional (Q1D) rare-earth chains, and (ii) their ferromagnetism are both of easy-plane anisotropy\cite{Vobwinkel-CeRh6Ge4,Matsuoka-CeRh6Ge4,Shen-CeRh6Ge4FMQCP}. Theoretically, an easy-plane anisotropy projects out an equal-spin pair and thus creates a triplet resonant valence bond (tRVB) state [$(|\uparrow\uparrow\rangle+|\downarrow\downarrow\rangle)/2$]\cite{Shen-CeRh6Ge4FMQCP}. Macroscopic entanglement can thus be injected into the ground state, i.e., easy-plane anisotropy in FM systems plays a similar role as that of magnetic frustration in antiferromagnetic systems\cite{Coleman-JLTP2010}, and this favors unconventional, local quantum criticality where the suppression of order parameter is also accompanied with a destruction of Kondo screening and a reconstruction of Fermi surface\cite{Custers-YbRh2Si2QCP,Schroder-CeCu6AuQCP,Paschen-YbRh2Si2Hall,Shishido-CeRhIn5dHvA,LuoY-CeNiAsOQCP,Komijani-FMQCP,Yamamoto-FM,Kirkpatrick-FMQCP,WangJ-FMQCP}. Moreover, as the tRVB states are already present, spin-triplet superconductivity is possible if they can be condensed. For progress, extensive FM materials with Q1D easy-plane anisotropy are needed to both testify this idea and further guide its development.

\begin{figure*}[!htp]
\vspace*{-40pt}
\hspace*{-10pt}
\includegraphics[width=16cm]{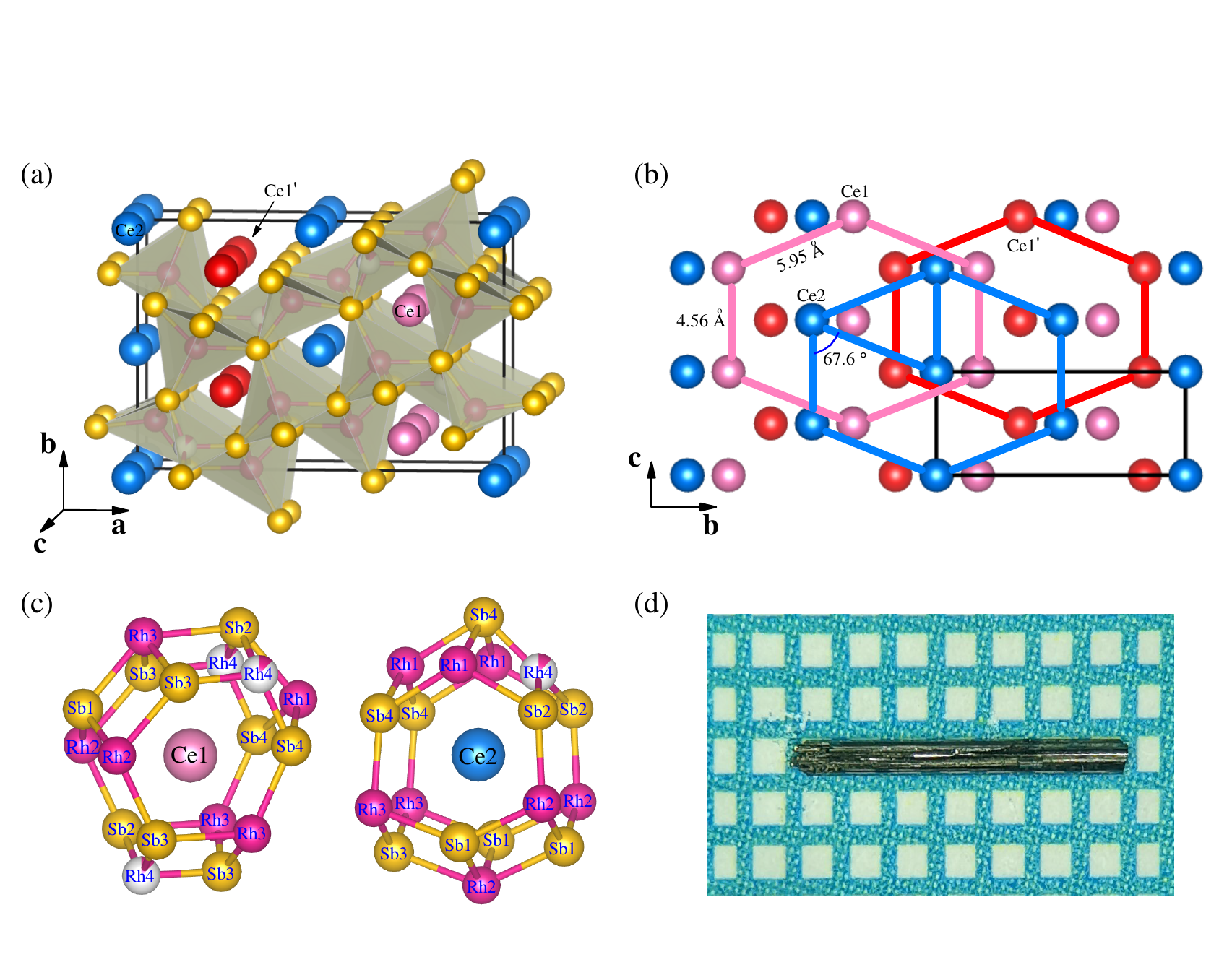}
\vspace*{-10pt}
\caption{\label{Fig1} (a) The crystalline structure of Ce$_2$Rh$_{3+\delta}$Sb$_4$. The Ce chains are along $\mathbf{c}$-axis, isolated by the Rh-Sb polyhedra. The partially occupied Rh4 is double-colored. The black rectangle depicts the unit cell. (b) Both Ce1 and Ce2 sub-lattices form distorted triangular networks within $\mathbf{bc}$ planes, stacked following the sequence of Ce2-Ce1'-Ce2-Ce1-Ce2 along $\mathbf{a}$. (c) Local environments of Ce1 and Ce2. (d) A photograph of Ce$_2$Rh$_{3+\delta}$Sb$_4$ single crystal on millimeter-grid paper. The single crystals are needle-like along $\mathbf{c}$ axis.}
\end{figure*}

Recently, we reported the growth of a new ternary superconducting rhodium-antimonide La$_2$Rh$_{3+\delta}$Sb$_4$ ($T_c\sim 0.8$ K)\cite{Cheng_La2Rh3Sb4}. Previously, this family of compounds (e.g. Ce$_2$Rh$_3$Sb$_4$) were recognized to crystallize in the orthorhombic Pr$_2$Ir$_3$Sb$_4$ structure, with the space group $Pnma$ (No.~62)\cite{GilR-Ce234,Schafer-Ce234}. By careful \emph{single-crystal} X-ray diffraction (XRD) and Scanning transmission electron microscope (STEM) measurements\cite{Cheng_La2Rh3Sb4}, we demonstrated the existence of an additional Rh4 site (that was not identified before) whose occupation rate is $\delta\approx 1/8$. In this compound, the La atoms form Q1D chains along $\mathbf{c}$ (Note \cite{note1}), and these chains are isolated by Rh-Sb polyhedra, which motivates us to further look into its sister compound Ce$_2$Rh$_{3+\delta}$Sb$_4$. In fact, as early as in 2011, Gil \textit{et al} already reported the growth of polycrystalline Ce$_2$Rh$_{3}$Sb$_4$\cite{GilR-Ce234}. There the temperature dependence of magnetic susceptibility down to 2 K remains Curie-Weiss-like without showing any trace of magnetic ordering. This implies that the system probably sits in a regime close to a quantum criticality. To better clarify the ground state and the possibility of QCP in this compound, both single-crystal samples and sub-Kelvin measurements are badly demanded.

In this paper, we report the synthesis and physical properties of Ce$_2$Rh$_{3+\delta}$Sb$_4$ ($\delta\approx 1/8$) single crystals. Our work manifests that Ce$_2$Rh$_{3+\delta}$Sb$_4$ is an easy-plane antiferromagnet with N\'{e}el temperature $T_N=1.4$ K. Short-range magnetic correlation can be observed well above $T_N$. Kondo scale is estimated, $T_K\sim2.4$ K, comparable to the strength of magnetic exchange. We also discuss the reason for the low ordering temperature, and the possibility of QCP under pressure.

\section{\Rmnum{2}. Experimental details}

Single crystalline Ce$_2$Rh$_{3+\delta}$Sb$_4$ was grown by a Bi-flux method as described in our earlier work\cite{Cheng_La2Rh3Sb4}. Ce chunk (Alfa Aesar, 99.9\%), Rh powder (Aladdin, 99.95\%), Sb granule (Aladdin, 99.9999\%), and Bi granule (Aladdin, 99.9999\%) were weighed in a molar ratio of
2:3:4:40, and transferred into an alumina crucible which was sealed in an evacuated quartz tube. The latter was heated to 500 $^{\circ}$C in 20 hours, held for one day, and then raised to 1100 $^{\circ}$C in 30 hours. After keeping at this temperature for 100 hours, we slowly cooled it to 500 $^{\circ}$C in two weeks, and then the Bi-flux was removed by centrifugation. The remaining Bi flux can be dissolved by a mixture of an equal volume of hydrogen peroxide and acetic acid. The as-grown samples mostly are needle-like with typical length 3-7 mm along the \textbf{c} axis, and about 0.6 $\times$$ $ 0.7$ $ mm$^2$ in cross section [Fig.~\ref{Fig1}(d)].

The chemical composition of the obtained single crystals was verified by energy dispersive x-ray spectroscopy (EDS) affiliated to a field emission scanning electron microscope (FEI Model SIRION), which gives the atomic ratio Ce : Rh : Sb = 23.21 : 32.56 : 44.24. Single crystalline x-ray diffraction (XRD) data were recorded at room temperature on a Rigaku XtaLAB mini II with Mo radiation ($\lambda_{K\alpha}$ = 0.71073 \AA). Electrical resistivity was measured by the standard four-probe method in a Physical Property Measurement System (PPMS-9, Quantum Design) which was also used for the specific heat measurements. In order to measure the in-plane resistivity, we employed focused-ion-beam (FIB) technique to make the microstructured device, as shown in the inset to Fig.~\ref{Fig2}(a). Magnetization measurements were performed using a Magnetic Property Measurement System (MPMS-VSM, Quantum Design) that is equipped with a $^3$He refrigerator.

\section{\Rmnum{3}. Results and Discussion}

The crystalline structure of Ce$_2$Rh$_{3+\delta}$Sb$_4$ determined from single-crystal XRD analysis is shown in Fig~\ref{Fig1}(a). The obtained cell parameters are: $a$=16.1927(11) \AA, $b$=4.5613(3) \AA, $c$=10.9917(8) \AA \cite{note1}, $\alpha$ = $\beta$ = $\gamma$ = 90 $^{\circ}$, and $Z$ = 4. The occupation rate of Rh4 is 0.123(3), close to that in La$_2$Rh$_{3+\delta}$Sb$_4$\cite{Cheng_La2Rh3Sb4}. Ce atoms form the chains along the $\mathbf{c}$ direction, and these chains are well separated by Rh-Sb polyhedra. It is worthwhile to mention that viewing along $\mathbf{a}$, the Ce atoms in each $\mathbf{bc}$ plane build up a distorted triangular network, and these networks are stacked along $\mathbf{a}$ following the sequence of Ce2-Ce1'-Ce2-Ce1-Ce2, seeing Fig.~\ref{Fig1}(b). Such a triangular sublattice is speculated to cause some frustration effect to Ce magnetism and will be further discussed later on. The local environments of Ce1 and Ce2 are depicted in Fig.~\ref{Fig1}(c). More details about the crystallline structure of Ce$_2$Rh$_{3+\delta}$Sb$_4$ can be found in Tables \ref{tbl:Structure} and \ref{tb2:Atoms}.

\begin{table}[!ht]
\tabcolsep 0pt \caption{\label{tbl:Structure} Single-crystal XRD refinement data for Ce$_2$Rh$_{3+\delta}$Sb$_4$ at 290 K.}
\vspace*{-15pt}
\begin{center}
\def\temptablewidth{1\columnwidth}
{\rule{\temptablewidth}{1pt}}
\begin{tabular*}{\temptablewidth}{@{\extracolsep{\fill}}cc}
composition ~~~~~~~~~~~~~~~~~~~~~~~~~~~~~~~~~~~         &  Ce$_2$Rh$_{3.123}$Sb$_4$           \\
formula weight (g/mol) ~~~~~~~~~~~~~~~~~~~~      &  1088.32                     \\
space group ~~~~~~~~~~~~~~~~~~~~~~~~~~~~~~~~~~~         &  $Pnma$ (No. 62)             \\
$a$ (\AA) ~~~~~~~~~~~~~~~~~~~~~~~~~~~~~~~~~~~~~~~~~~~           &  16.1927(11)     \\
$b$ (\AA)  ~~~~~~~~~~~~~~~~~~~~~~~~~~~~~~~~~~~~~~~~~~~          &  10.9917(8)   \\
$c$ (\AA) ~~~~~~~~~~~~~~~~~~~~~~~~~~~~~~~~~~~~~~~~~~~           &  4.5613(3)      \\
$V$ (\AA$^{3}$) ~~~~~~~~~~~~~~~~~~~~~~~~~~~~~~~~~~~~~~~~~~     &  811.84(10)    \\
$Z$  ~~~~~~~~~~~~~~~~~~~~~~~~~~~~~~~~~~~~~~~~~~~~~~~~~ &  4               \\
$\rho$ (g/cm$^{3}$) ~~~~~~~~~~~~~~~~~~~~~~~~~~~~~~~~~~~~~~ &  8.904                     \\
$2\theta$ range ~~~~~~~~~~~~~~~~~~~~~~~~~~~~~~~~~~~~~~~~     &  4.478-53.996$^{\circ}$         \\
no. of reflections, $R_{int}$ ~~~~~~~~~~~~~~~~~~~~~~  &   9097, 0.0749                           \\
no. of independent reflections ~~~~~~~~~~~~ &   959                                    \\
no. of parameters ~~~~~~~~~~~~~~~~~~~~~~~~~~~ &   62                  \\
$R_1^{\dag}$, $w R_2^{\ddag}$ [$I>2\delta(I)$] ~~~~~~~~~~~~~~~~~~~~~~~~~~~&   0.0282, 0.0578    \\
$R_1$, $w R_2$ (all data) ~~~~~~~~~~~~~~~~~~~~~~~~~~~ &   0.0307, 0.0587          \\
goodness of fit on $F^2$ ~~~~~~~~~~~~~~~~~~~~~~~ & 1.108                                           \\
largest diffraction peak and hole (e/\AA$^{3}$)   &   2.02 and $-$1.90       \\
\end{tabular*}
{\rule{\temptablewidth}{1pt}}
\end{center}
$^{\dag}$ $R_1=\sum||F_{obs}|-|F_{cal}||/\sum|F_{obs}|$. ~~~~~~~~~~~~~~~~~~~~~~~~~~~~\\
$^{\ddag}$ $w R_2=[\sum w(F_{obs}^2-F_{cal}^2)^2/\sum w (F_{obs}^2)^2]^{1/2}$, ~~~~~~~~~~~~~~~\\
$w=1/[\sigma^2F_{obs}^2+(a\cdot P)^2+b\cdot P]$,~~~~~~~~~~~~~~~~~~~~~~~~~~\\
where $P=[\mathrm{max}(F_{obs}^2)+2F_{cal}^2]/3$. ~~~~~~~~~~~~~~~~~~~~~~~\\
\end{table}

\begin{table*}[!ht]
 \caption{\label{tb2:Atoms} Structural parameters and equivalent isotropic displacement parameters $U_{eq}$ of Ce$_2$Rh$_{3+\delta}$Sb$_4$. $U_{eq}$ is taken as 1/3 of the trace of the orthogonalized $U_{ij}$ tensor.}
\vspace*{0pt}
\centering
\setlength{\tabcolsep}{6mm}
\begin{tabular*}{0.9\linewidth}{@{}ccccccc@{}}
\hline
 Atoms   &  Wyck.  &  $x$         &     $y$     &     $z$      &     Occ.      & $U_{eq}$ (\AA$^2$) \\\hline
  Ce1    &  4c     &  0.25047(4)  &  0.91519(6) &  0.2500      &    1.000      &  0.00609(17)       \\
  Ce2    &  4c     &  0.50390(4)  &  0.74876(6) &  0.2500      &    1.000      &  0.00874(17)       \\
  Sb1    &  4c     &  0.43235(4)  &  0.05274(6) &  0.2500      &    1.000      &  0.00581(19)         \\
  Sb2    &  4c     &  0.65227(4)  &  0.27291(6) &  0.2500      &    1.000      &  0.00750(19)         \\
  Sb3    &  4c     &  0.29111(5)  &  0.11744(7) &  0.7500      &    1.000      &  0.0079(2)         \\
  Sb4    &  4c     &  0.39686(5)  &  0.43626(7) &  0.2500      &    1.000      &  0.00861(19)         \\
  Rh1    &  4c     &  0.55703(5)  &  0.46494(8) &  0.2500      &    1.000      &  0.0052(2)         \\
  Rh2    &  4c     &  0.59242(5)  &  0.05378(8) &  0.2500      &    1.000      &  0.0065(2)         \\
  Rh3    &  4c     &  0.31887(5)  &  0.22375(8) &  0.2500      &    1.000      &  0.0070(2)         \\
  Rh4    &  4c     &  0.3485(7)   &  0.6227(9)  &  0.2500      &    0.123(3)   &  0.029(2)        \\\hline
\end{tabular*}
\end{table*}

\begin{figure}[!ht]
\vspace*{-0pt}
\hspace*{-0pt}
\includegraphics[width=9cm]{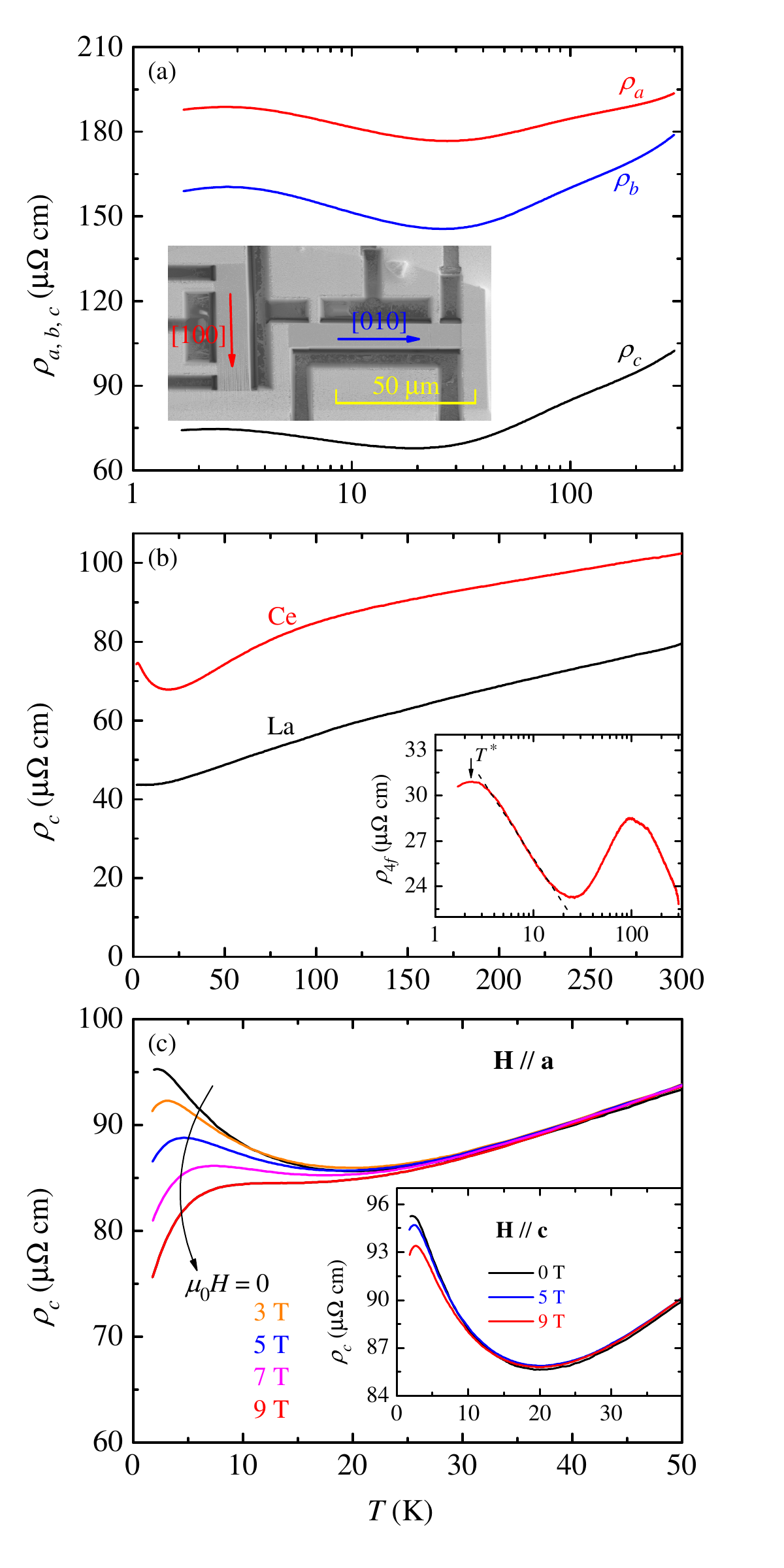}
\vspace*{-15pt}
\caption{\label{Fig2} (a) $\rho_{a}$, $\rho_{b}$ and $\rho_c$ as functions of $T$ for Ce$_2$Rh$_{3+\delta}$Sb$_4$. The inset shows the scanning-electron-microscope image of the microstructured device made by FIB to measure $\rho_{a}$ and $\rho_{b}$. (b) Temperature dependent $\rho_c$ of $Ln_2$Rh$_{3+\delta}$Sb$_4$ [$Ln$=La (black) and Ce (red)]. The inset shows $\rho_{4f}\equiv \rho_c^{Ce}-\rho_c^{La}$. $T^*$ signifies the temperature where $\rho_{4f}$ starts to decrease. (c) $\rho_{c}(T)$ profiles at different magnetic fields $\mathbf{H}\parallel\mathbf{a}$. The inset shows the results for $\mathbf{H}\parallel\mathbf{c}$.}
\end{figure}

Temperature dependent resistivity of Ce$_2$Rh$_{3+\delta}$Sb$_4$ for electrical current parallel to different principal axes are shown in Fig.~\ref{Fig2}(a). The base temperature of our resistivity measurements was 1.7 K. The profiles of $\rho_{a,b,c}(T)$ look similar except for their different magnitudes, $\rho_a>\rho_b>\rho_c$, following the same sequence of $a$, $b$, and $c$ lengths. We should also point out that the inter-chain resistivity $\rho_{a,b}$ is about two times of that of the intra-chain $\rho_{c}$, characteristic of quasi-1D electronic property. At low temperature below $\sim 25$ K, a $-\log{T}$ dependence can be observed in all $\rho_{a,b,c}(T)$, which is commonly associated with the incoherent Kondo effect in heavy-fermion materials.

To better understand the electronic correlation effect in  Ce$_2$Rh$_{3+\delta}$Sb$_4$, we now focus on $\rho_{c}$; for comparison, the results of La$_2$Rh$_{3+\delta}$Sb$_4$ are also displayed, seeing Fig.~\ref{Fig2}(b). While La$_2$Rh$_{3+\delta}$Sb$_4$ behaves as a conventional metal, the $\rho_c(T)$ of Ce$_2$Rh$_{3+\delta}$Sb$_4$ shows rather different features. Upon cooling, $\rho_c(T)$ decreases linearly until $\sim$ 100 K where its slope gradually increases. Below $\sim$ 25 K, $\rho_c$ turns up again. To get a clear insight into the scattering rate due to the $4f$ electrons, we calculate $\rho_{4f}$ by subtracting $\rho_c^{La}$ from $\rho_c^{Ce}$, as shown in the inset to Fig.~\ref{Fig1}(b). The bump centered around 100 K now can be clearly seen in $\rho_{4f}$. We attribute this to spin scatterings caused by the excitations between crystalline electric field (CEF) levels, and this can be further supported by specific heat measurements (see below). Incoherent Kondo scattering sets in below 25 K, as evidenced by the logarithmic $T$ dependent $\rho_{4f}$. At low temperature, $\rho_{4f}$ maximizes at $T^*=2.4$ K below which it falls back. At a first glance, such a peak in $\rho_{4f}(T)$ might be reminiscent of the onset of Kondo coherence, however, we notice that this peak is strongly field-dependent, seeing Fig.~\ref{Fig2}(c). This is especially so when magnetic field is applied within the easy plane, viz, the peak at $T^*$ moves upwards rapidly and meanwhile its height is also suppressed drastically. Such a significant field dependence is far beyond a regular field effect on coherent Kondo scattering\cite{Christianson-CeRhIn5MR}, but is more attributable to a field-induced polarization that reduces spin-flip scattering\cite{Krellner-CeRuPO,LuoY-CeNi2As2}. In this sense, magnetic ordering or at least short-range magnetic correlation would be likely at lower temperature. To clarify this issue, we turn to magnetic susceptibility measurements.

\begin{figure}[!ht]
\vspace*{-15pt}
\hspace*{-20pt}
\includegraphics[width=9cm]{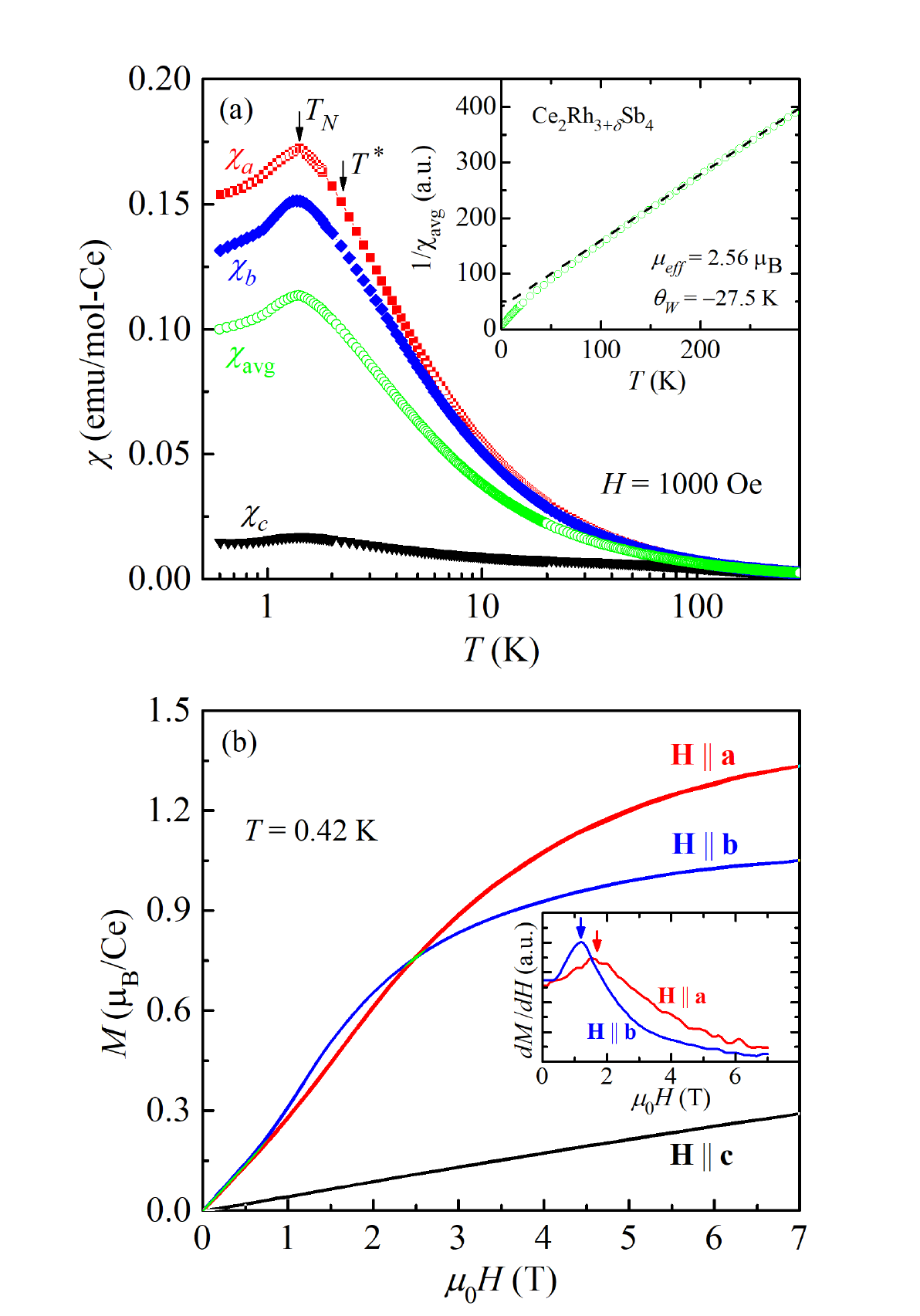}
\vspace*{-0pt}
\caption{\label{Fig3}(a) Temperature dependence of the magnetic susceptibility $\chi$ at 0.1 T measured with different field orientations. The inset shows $1/\chi_{avg}$ as a function of $T$, where $\chi_{avg}$ is the powder-averaged magnetic susceptibility. The dashed line stands for Curie-Weiss fitting. (b) Isothermal field dependent magnetization at 0.42 K. The inset of (b), $dM/dH$ curves. The red and blue arrows mark the metamagnetic transitions observed in $M_a$ and $M_b$, respectively.}
\end{figure}

Figure \ref{Fig3}(a) shows the temperature dependence of magnetic susceptibility ($\chi\equiv M/H$) of Ce$_2$Rh$_{3+\delta}$Sb$_4$, measured for different field orientations. One clearly sees that $\chi_a$ and $\chi_b$ are much larger than $\chi_c$, confirming $\mathbf{ab}$ as the easy plane. The powder-averaged magnetic susceptibility is calculated via $\chi_{avg}=(\chi_a+\chi_b+\chi_c)/3$. For temperature above 100 K, $\chi_{avg}(T)$ obeys the standard Curie-Weiss formula, $\chi_{avg}(T)$=$C_0/(T-\theta_W)$, where $C_0$ is the Curie constant and $\theta_W$ is the Weiss temperature (Note that the temperature independent term $\chi_0$ turns out to be negligibly small here). The fitting yields $\theta_W=-27.5$ K and the effective moment $\mu_{eff}=2.56~\mu_B$, very close to that of a free Ce$^{3+}$ ion, 2.54 $\mu_B$. This implies that the $4f$ electron is highly localized and Rh ions are essentially nonmagnetic. The derived negative $\theta_W$ suggests antiferromagnetic correlation dominant in Ce$_2$Rh$_{3+\delta}$Sb$_4$. Indeed, $\chi_a$, $\chi_b$ and $\chi_c$ all show a clear peak near $T_N=1.4$ K, characteristic of AFM transition.

The AFM ordering in Ce$_2$Rh$_{3+\delta}$Sb$_4$ is further supported by isothermal field dependent magnetization [$M(H)$] shown in Fig.~\ref{Fig3}(b). At 0.42 K, well below $T_N$, $M_c$ increases linearly with $H$, and reaches 0.3 $\mu_B$/Ce at 7 T; whereas $M_a$ and $M_b$ grow rapidly with $H$ and tend to saturate at 1.4 and 1.1 $\mu_B$/Ce, respectively. Such an anisotropy reaffirms that $\mathbf{c}$ is the hard axis, and the Ce moments are aligned within $\mathbf{ab}$ plane. Another important feature in $M(H)$ is that the slope maximizes at 1.7 and 1.2 T for $M_a$ and $M_b$, respectively, which may correspond to field-induced metamagnetic transitions. This provides additional evidence for the AFM ground state in Ce$_2$Rh$_{3+\delta}$Sb$_4$.

\begin{figure}[!ht]
\vspace*{-0pt}
\hspace*{-5pt}
\includegraphics[width=9cm]{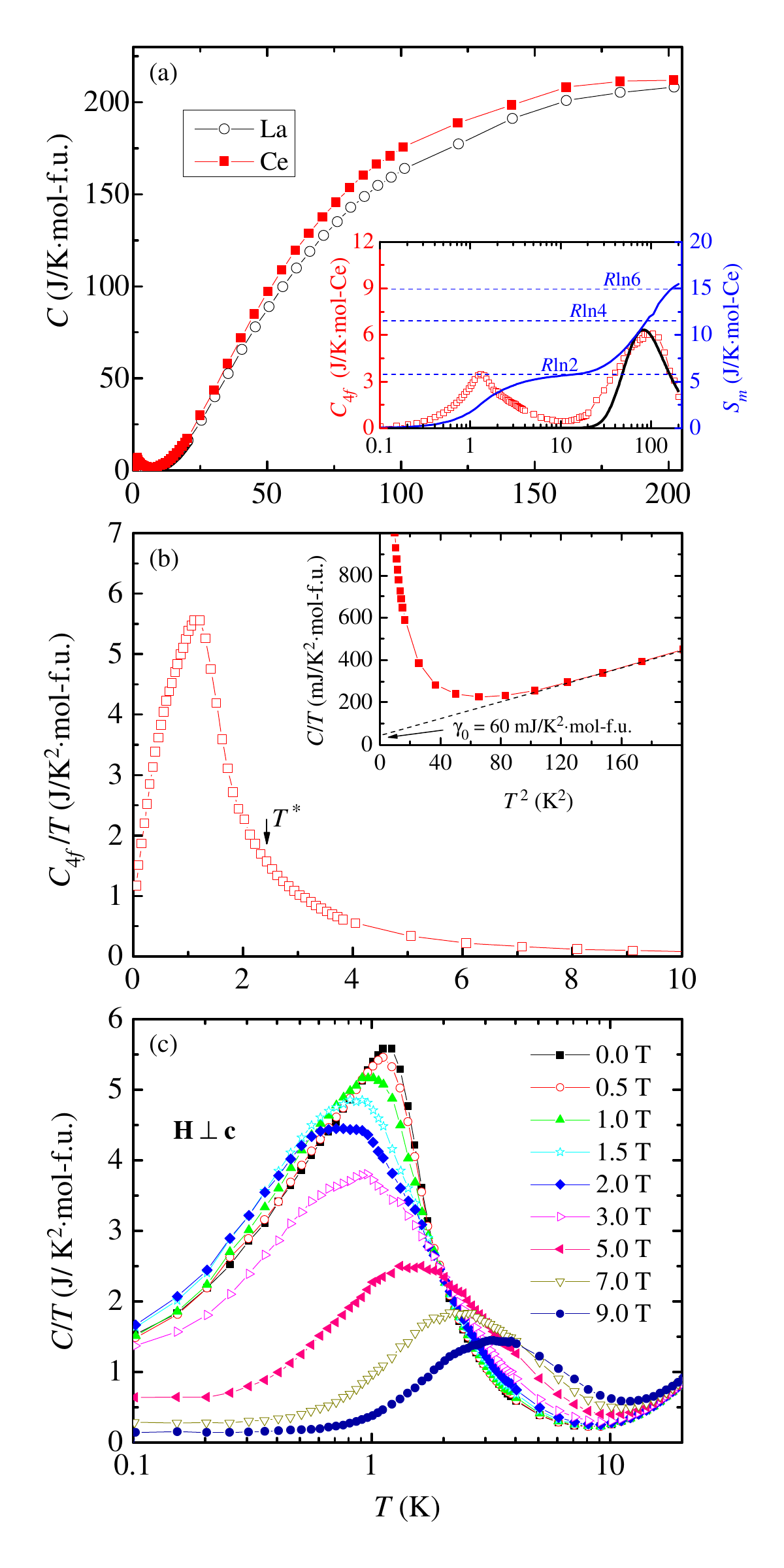}
\vspace*{-20pt}
\caption{\label{Fig4}  (a) Temperature dependent specific heat of Ce$_2$Rh$_{3+\delta}$Sb$_4$ and La$_2$Rh$_{3+\delta}$Sb$_4$. The inset shows the specific heat from $4f$ electrons and magnetic entropy (per molar Ce) as functions of $T$. The black solid line is a 3-doublet Schottky anomaly fitting with $\Delta_1=210$ K and $\Delta_2=227$ K. (b) $C_{4f}/T$ vs. $T$. Note that the nuclear quadrupolar Schottky contribution that appears below 0.3 K has already been removed by properly fitting to a $1/T^3$ law\cite{Gopal-SpeciHeat}. The inset plots $C/T$ vs. $T^2$ which yields $\gamma_0=60$ mJ/K$^2\cdot$mol-f.u. (c) Temperature dependence of $C/T$ measured at various magnetic fields $\mathbf{H}\perp\mathbf{c}$.}
\end{figure}

In Fig.~\ref{Fig4}(a), we present the temperature dependent specific heat ($C$) of  Ce$_2$Rh$_{3+\delta}$Sb$_4$. For comparison, $C(T)$ of the non-$f$ reference La$_2$Rh$_{3+\delta}$Sb$_4$ is also shown. At high temperature, $C(T)$ tends to saturate to the classic limit $3NR=224.5$ J/K$\cdot$mol-f.u. given by Dulong-Petit's law\cite{Ashcroft-SSP}, where $R=8.314$ J/K$\cdot$mol is the ideal gas constant, and $N=9$ is the number of atoms in a formula unit of stoichiometric Ce$_2$Rh$_3$Sb$_4$. A peak can be seen in $C(T)$ at around 1.4 K, which should be a consequence of the magnetic phase transition as observed in magnetic susceptibility. The normal-state Sommerfeld coefficient can be derived by linearly extrapolating $C/T$ vs. $T^2$ in the paramagnetic regime to zero, $\gamma_0=60$ mJ/K$^2\cdot$mol-f.u. [cf inset to Fig.~\ref{Fig4}(b)], a value moderately enhanced by a factor of 3 as compared to that of La$_2$Rh$_{3+\delta}$Sb$_4$\cite{Cheng_La2Rh3Sb4}. Such an enhancement should arises from 4$f$ electronic correlation effect. The contribution from $4f$ electrons, $C_{4f}$, is derived by subtracting the specific heat of La$_2$Rh$_{3+\delta}$Sb$_4$, and the resultant is displayed in the inset to Fig.~\ref{Fig4}(a). A broad peak centered at $\sim$100 K is clearly visible, which should be a consequence of Schottky anomaly due to CEF splitting of Ce$^{3+}$ multiplet ($J=5/2$).


The low-temperature AFM transition is now clearly seen in $C_{4f}/T$ as shown in the main frame of Fig.~\ref{Fig4}(b). It should be pointed out that this peak does not appear $\lambda$-shaped, but is rather symmetric and has a long tail at the right-hand side, suggesting the presence of short-range ordering above $T_N$ and is in agreement with the features of $\rho_c(T)$ and $\chi(T)$.

We then investigate the magnetic entropy $S_m$ (per molar cerium) by integrating $C_{4f}/T$ over $T$, as shown in the inset to Fig.~\ref{Fig4}(a). The entropy gain reaches 0.49$R\ln2$ and 0.73$R\ln2$ at $T_N$ and $T^*$, respectively, manifesting a Kramers doublet ground state of Ce$^{3+}$ in CEF. $S_m$ gets fully recovered $R\ln2$ near 10 K. The Kondo scale ($T_K$) can be estimated from $S_m(T_K/2)=0.4R\ln2$ \cite{Gegenwart2008}. This analysis yields $T_K\sim2.4$ K, which is nearly the same as $T^*$. Therefore, Ce$_2$Rh$_{3+\delta}$Sb$_4$ represents a Kondo-lattice compound whose Kondo scale is comparable to the magnetic exchange, and furthermore, short-range ordering has played an important role to the reduced entropy gain at $T_N$. $S_m$ approaches $R\ln6$ at around 200 K, manifesting that the system has gradually recovered the full sextet degeneracy of Ce$^{3+}$. This is demonstrated by a 3-doublet Schottky anomaly fitting with $\Delta_1=210$ K and $\Delta_2=227$ K, where $\Delta_{i}$ ($i$=1,2) are the energy difference between the ground and the $i$th excited doublet.

We also notice a salient feature that $C_{4f}/T$ remains as large as $\sim1.2$ J/K$^2\cdot$mol-f.u. even at the ultra-low temperature 0.05 K. Such a large residual specific heat coefficient is rather unusual for a long-range ordered antiferromagnet. In addition, a slope change is seen near 0.6 K in $C_{4f}/T$ [Fig.~\ref{Fig4}(b)]. A naive thinking is that a spin-reorientation takes place gradually within the AFM state. However, considering the two inequivalent Ce sites in the crystal structure, an alternative possibility could be that one set of the two Ce sub-lattices gives rise to the AFM transition at $T_N$, while the other Ce sublattice remains disordered (or becomes ordered at a lower temperature). This gives a reasonable explanation to the large residual specific heat coefficient at $T\rightarrow 0$, as well as the slope change in $C_{4f}/T$.

In fact, the two inequivalent Ce sublattices can be also reflected in specific heat under magnetic field ($\mathbf{H}\perp\mathbf{c}$), as shown in Fig.~\ref{Fig4}(c). On the whole, for a small applied field, the specific heat peak near $T_N$ is suppressed, consistent with the nature of AFM transition; further increasing field, the peak gets broader and shifts to higher temperatures. Such an evolution is the consequence of the Schottky anomaly due to the reopening of magnon gap caused by Zeeman splitting. This reflects a field-induced ferromagnetic ordering by undergoing a metamagnetic transition and is in agreement with $\rho_c(T, H)$ and $M(H)$ aforementioned.Meanwhile, we notice a shoulder appears in $C/T$ curve, which is most obviously seen at 2.0 and 3.0 T [Fig.~\ref{Fig4}(c)]. The appearance of this shoulder is probably owing to the response to external field from the originally disordered Ce sublattice. For larger fields, this shoulder gradually merges into the main peak and is no longer resolved, because both Ce sublattices are getting polarized. However, we must admit that to get a precise view of the magnetic structure, experiments like neutron scattering are required.

Finally, several additional remarks are in order:

(1) The origin of the low $T_N$. There could be multiple reasons for the low $T_N$ in Ce$_2$Rh$_{3+\delta}$Sb$_4$. First, within a single Ce sub-lattice, the Ce-Ce distances are 4.56 \AA ~and 5.95 \AA ~[Fig.~\ref{Fig1}(b)], both of which are much larger than the typical values (3.8-4.2 \AA) in most Ce-based intermetallic compounds\cite{Rai-CeIr3Ge7}. The large Ce-Ce distance weakens the magnetic exchange severely. Note that the Ruderman-Kittel-Kasuya-Yosida (RKKY) exchange coupling decays rapidly with the intersite distance ($r_{ij}$) following $J_{RKKY}\propto\frac{x\cos{x}-\sin{x}}{x^4}$, where $x=2k_F r_{ij}$ and $k_F$ is Fermi wavevector\cite{RKKY-RK}. The same scenario was also adopted for the low ordering temperature in Ce$_3$Pt$_{23}$Si$_{11}$\cite{Opagiste-Ce3Pt23Si11}, Ce$Tm_2$Cd$_{20}$ ($Tm$=Co,Ni)\cite{White-CeT2Cd20} and CeIr$_3$Ge$_7$\cite{Rai-CeIr3Ge7}. In addition, since each Ce sub-lattice form a distorted triangular network as displayed in Fig.~\ref{Fig1}(b), geometric frustration probably will further reduce the ordering temperature. The low $T_N$ in Ce$_2$Rh$_{3+\delta}$Sb$_4$, as well as the comparable magnetic correlation and Kondo scales, makes it a novel candidate for further investigating quantum criticality under pressure.

(2) The two inequivalent Ce sites. Although Ce1- and Ce2-sublattices build up the same distorted triangular networks [cf Fig.~\ref{Fig1}(b)], their local environments are different. As shown in Fig.~\ref{Fig1}(c), both Ce1 and Ce2 are surrounded by 9 Rh (including Rh4) and 9 Sb sites, but Ce1 has three Rh4 neighbours, while Ce2 only has one. It is reasonable to infer that there are more conduction electrons around Ce2, correspondingly, the extent of Kondo screening should also be stronger at Ce2. In this sense, strictly speaking, one would expect two different Kondo scales in this compound. 
It is interesting to mention that multiple Ce-sites with different strengths of Kondo coupling and correspondingly distinct ground states have also been observed in Ce$_2$Rh$_3$Sn$_5$\cite{Gamza-Ce2Rh3Sn5} and Ce$_3$Rh$_4$Sn$_7$\cite{Opletal-Ce3Rh4Sn7}.

\section{\Rmnum{4}. Conclusions}

To summarize, by a Bi-flux method, we successfully synthesized single crystalline Ce$_2$Rh$_{3+\delta}$Sb$_4$, a new dense Kondo lattice material. The compound contains quasi-1D Ce chains along $\mathbf{c}$ which is also the magnetic hard axis. Quasi-1D electronic and magnetic features are unveiled by resistivity and magnetic susceptibility measurements. A long-range antiferromagnetic transition occurs at $T_N=1.4$ K, while clear short-range ordering can be detected well above $T_N$. We attribute the low ordering temperature to the large Ce-Ce distance as well as the geometric frustration. Kondo scale is estimated to be about 2.4 K, comparable to the strength of magnetic exchange. Our work provides a rare paradigm of dense Kondo lattice whose RKKY exchange and Kondo coupling are both weak but competing.


\section{Acknowledgments}

The authors acknowledge Joe D. Thompson and Jinke Bao for helpful discussions. This work is supported by National Key R\&D Program of China (2022YFA1602602 and 2022YFA1402200), the open research fund of Songshan Lake Materials Laboratory (2022SLABFN27), National Natural Science Foundation of China (11974306 and 12034017), Guangdong Basic and Applied Basic Research Foundation (2022B1515120020), and Key R\&D Program of Zhejiang Province, China (2021C01002).




\begin{thebibliography}{40}
\bibitem [{not()}]{note1}%
  \BibitemOpen
  \href@noop {} {}\bibinfo {note} {In this work, we have exchanged the
  $\mathbf{b}$ and $\mathbf{c}$ axes as compared with those in
  Ref.~\cite{Cheng_La2Rh3Sb4}.}\BibitemShut {Stop}%
\end{thebibliography}

%

\end{document}